# Nano1D: An accurate Computer Vision software for analysis and segmentation of low-dimensional nanostructures

Ehsan Moradpur-Tari*[a], Sergei Vlassov[a,b], Sven Oras[a,b], Mart Ernits[a], Elyad Damerchi[a], Boris Polyakov[c], Andreas Kyritsakis[a], and Veronika Zadin[a]

[a]Institute of Technology, University of Tartu, Nooruse 1, 50411 Tartu, Estonia

[b]Institute of Physics, University of Tartu, W. Ostwaldi 1, 50411 Tartu, Estonia

[c]Institute of Solid State Physics, University of Latvia, Kengaraga street 8, LV-1063 Riga, Latvia

*Corresponding Author: ehsan.moradpur.tari@ut.ee

## Abstract

Nanoparticles in microscopy images are usually analyzed qualitatively or manually and there is a need for autonomous quantitative analysis of these objects. In this paper, we present a physics-based computational model for accurate segmentation and geometrical analysis of one-dimensional deformable overlapping objects from microscopy images. This model, named Nano1D, has four steps of preprocessing, segmentation, separating overlapped objects and geometrical measurements. The model is tested on SEM images of Ag and Au nanowire taken from different microscopes, and thermally fragmented Ag nanowires transformed into nanoparticles with different lengths, diameters, and population densities. It successfully segments and analyzes their geometrical characteristics including lengths and average diameter. The function of the algorithm is not undermined by the size, number, density, orientation and overlapping of objects in images. The main strength of the model is shown to be its ability to segment and analyze overlapping objects successfully with more than 99% accuracy, while current machine learning and computational models suffer from inaccuracy and inability to segment overlapping objects. Benefiting from a graphical user interface, Nano1D can analyze 1D nanoparticles including nanowires, nanotubes, nanorods in addition to other 1D features of microstructures like microcracks, dislocations etc.

Graphical abstract

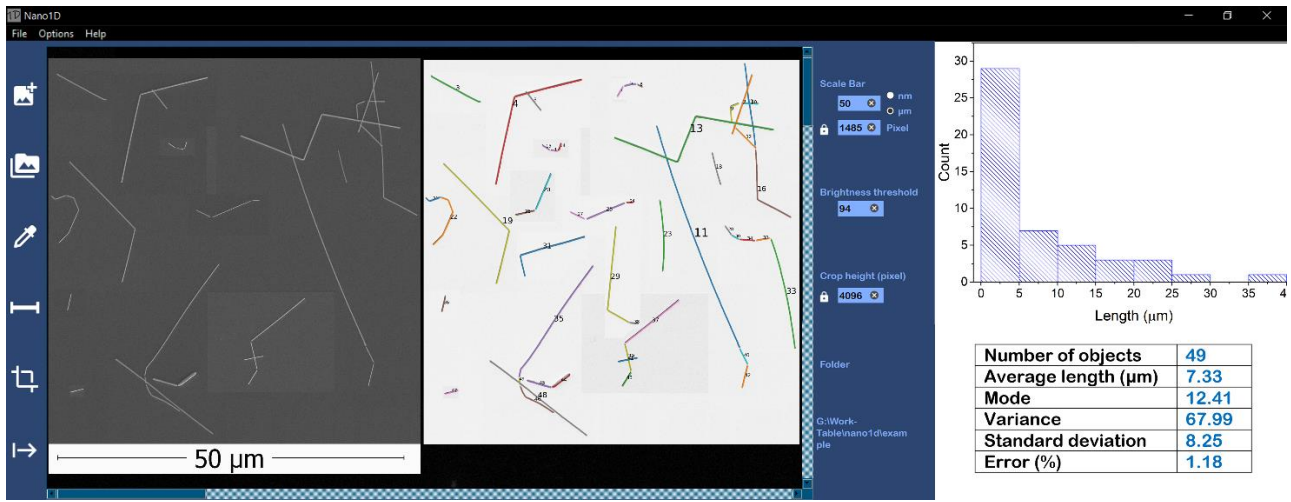

1  Introduction

With the advent of big data materials science, manual image analysis using related computer programs is not fast enough and there is a need for automation. While autonomous design and synthesis has attracted some attention [1–3], autonomous characterization methods are still not sufficiently versatile and reliable. Microscopy images are usually hard to analyze quantitatively and systematically, and there is a necessity for the development of efficient and accurate models to evaluate them. The rate of Electron Microscopy data acquisition has reached 200 Terabytes per hour in recent years and current efforts of autonomous analysis are more focused on object detection and segmentation of 2D objects in microscopy images [4].

Segmentation of 1D objects from the background using computer vision can be harder than higher dimensional objects [5] because of their thinner morphology. Nanowires and nanotubes, which are the two main one-dimensional (1D) structures in materials science [6], are amongst the most studied structures in electron microscopy. Machine learning (ML) has been commonly used for modeling physical properties of these materials [7]. The common approach in these studies uses artificial neural networks to model geometrical, mechanical, thermal, electrical, and electronic properties based on experimental or atomistic simulation data. Although training and prediction of ML models is fast, data preparation, labeling and annotation can be so demanding and needs a lot of human effort and precision. Moreover, the reliability of ML models is limited, and their accuracy can be an issue for measurements. This limited reliability can be catastrophic when we want to extend the application to load-bearing engineering structures. For example, when using ML for crack detection and propagation [8], the inaccuracy in detecting cracks that surpass the critical length can lead to serious failures. Therefore, efficient and precise surrogate models can be seen as firm grounds for further developments.

Computer Vision methods are usually based on ML rather than physical or computational methods and are more focused on classification and detection of nanoparticles [9,10], while measurements need segmentation of the objects from background. Lin et al [11] developed a deep learning model for nanowire segmentation, which could segment them with about 90% detection accuracy. Taher et al [1] developed a deep learning model for the classification of carbon nanotube forests, which could label the classes with 91% accuracy. The input data in both studies mentioned [1,11] were produced by algorithms rather than real microscopy images. Self-occlusion may also be seen in deformable objects and has been a subject of studies in the past decades [12]. Evan [13] developed a sweep-and-prune algorithm for detecting self-collision of cables and protein chains. Caporali et al [14] developed an approach for segmentation of deformable linear objects. In their case, a neural network was used for semantic segmentation, and overlapped wires were seen as one object. Then the instances were skeletonized and segmented into separate parts from intersections regardless of their overlapping. Afterwards, another neural network was trained to stitch various parts and to separate the wires. This method can segment overlapping wires with about 85% success, yet it needs multiple processing, annotating, and training steps. Nang [15] has also developed a 1D object segmentation method which is suitable for texture image recognition and classification purposes.

Agglomeration of nanostructures is inevitable under realistic conditions. This issue causes the overlapping of the objects in microscopy images. Self-collision of deformable elongated structures can also result in self-occlusion and measurement hindering. This makes measurement tasks more arduous for humans and even artificial intelligence (AI). Here, we developed an algorithm for successful and precise segmentation and geometrical measurements of 1D objects, which can address overlapping, agglomeration, and self-collision successfully. This algorithm uses human-mimicking logic for following and recognizing the trail of sections.

## 1 Materials and methods

### 1.1 Microscopy sample preparation and imaging

Samples for analysis were prepared by depositing silver nanowires (*BlueNano*) from ethanol solution onto oxidized silicon wafer (*Semiconductor Wafer, Inc.*). Fragmentation of Ag nanowires was achieved by thermal annealing in a muffle furnace (Nabertherm) in air atmosphere with various temperatures used for different samples. Synthesis of gold nanowires described elsewhere [16]. Samples were characterized by SEM (FEI Nova NanoSEM 450). Uncompressed SEM images of nanowires were taken at magnification 1500X with image dimensions of 6144×4415 pixels at 113 dpi and 24 bits. This allowed to capture tens of nanowires in a single image with sufficient resolution, so that the width of individual nanowires in the image had at least several (>3) pixels. Furthermore, gold nanowires were prepared similarly and imaged by another microscope (LYRA3 TESCAN) with magnification of 101kX.

1.2 Model architecture and packages

The model has been built in *Python 3.11.2* environment. Open-source python libraries of *scikit-image v0.20.0* and *SciPy 1.9.2* have also been used for image processing.

2 Theory

2.1 Image pre-processing

Depending on the samples investigated, microscopy images may require some manual editing and sorting before processing with algorithm. For instance, in addition to the structures of interest, microscopy images may have extra objects such as random dirt particles or some features (defects, patterns etc.) on the substrate. Moreover, some loosely bound structures can move under electron beam irradiation during imaging due to charging, creating extra features on the image. Such objects may introduce errors and therefore should be removed manually or the whole image should be sorted out if extra objects interfere with the measurements. Microscope users should also ensure proper contrast between objects of interest and background by choosing the right substrate and imaging parameters.

After reading the image, a Gaussian blur 3×3 kernel and standard deviation of 10 is applied to smooth the image and remove internal noises. Afterward, the edges can be sharpened by subtracting the smoothened image and normalized again to be between 0-255. These steps can be avoided based on the internal noises and type of the image and are embedded as optional in the code. A threshold is used to binarize the image which is selected manually. This threshold is the brightness of edge of objects and depending on the image's brightness is usually between 40-120. In the next step, the images are skeletonized using Zhang [17] or Lee [18] algorithms.

2.2 Finding sections, tails, and intersections

Object labeling is done with the *label* function of *scikit-image* with *connectivity=2*, which uses diagonal and orthogonal hops for connecting pixels. The main limitation of this function is that it cannot segment overlapping objects and label them as one object which leads to inaccurate measurements. This issue necessitated the development of the Nano1D algorithm.

To separate overlapping nanowires, first a one-pixel border with value of 0 is added to the image and then the image is filtered using a 3×3 kernel, called connectivity kernel ($\omega_c$), which value of its center element is 6 and other elements are equal to 1:

$$g(x,y) = \omega_c * f(x,y) = \sum_{i=-1}^{1} \sum_{j=-1}^{1} \omega_c(i,j) f(x-i, y-j) \qquad (1)$$

Where $g(x,y)$ is the filtered image and $f(x,y)$ is the original image. Then the nearest neighbor number, $N(x, y)$, of each pixel is calculated using the convolved image based on the following equation:

$$N(x,y) = (g(x,y) \times b(x,y) - 6) \times b'(x,y) \qquad (2)$$

Where $b(x,y)$ and $b'(x,y)$ are binary functions:

$$b(x,y) = \begin{cases} 1 & if\ g(x,y) > 6 \\ 0 & otherwise \end{cases} \tag{3}$$

$$b'(x,y) = \begin{cases} 1 & if\ g(x,y) \times b(x,y) > 0 \\ 0 & otherwise \end{cases} \tag{4}$$

After this step, segmentation starts from tails ($N(x,y)=1$) and the next connected pixel is found using a scoring matrix, $S_t(x,y)$, calculated by convolution of the connectivity kernel on neighbor pixels:

$$S_t(x,y) = \omega_c * N(x,y) - \omega_p = \left( \sum_{i=-1}^{1} \sum_{j=-1}^{1} \omega_c(i,j) N(x-i, y-j) \right) - \omega_p \tag{5}$$

Where $\omega_p$ is a 3×3 kernel with 6 in the center and 0 in other elements. So far, the scanning has taken its first step from the tail, but for the next steps it must use a more advanced score matrix to avoid returning to the first point. To achieve this, we must employ an orienteering matrix:

$$\theta(x,y) = \begin{bmatrix} -(x+y) & -y & x-y \\ -x & 6 & x \\ y-x & y & x+y \end{bmatrix} \tag{6}$$

The score matrix for the next steps is calculated by:

$$S_m(x,y) = \left( (\omega_c * f(x,y)) + 3 \times \theta(x,y) \right) \times f'(x,y) \tag{7}$$

Where $f'(x,y)$ is a 3×3 binary image with the current pixel in the center. This process continues until it reaches the next tail. So far, it can successfully segment and analyze separate 1D objects. However, when objects are overlapped, it cannot separate them well. To separate overlapping objects correctly, we needed to add a key step, which is described in the next section.

## 2.3 Separating overlapping objects

The main idea of this process is using the orientation of overlapping objects at their intersections to separate them. The logic of this process mimics the human's brain to follow trails at intersections. When reaching an intersection, the algorithm calculates the orientation of the last pixels of scanned object and composes a recall vector. Then evaluates the next options based on the orientation of their next pixels. After quantifying the options as vectors, the decision is made based on the dot product of option vectors and recall vector. The vectors are normalized with respect to their length. The number of pixels in each vector is 30 by default, but it can be changed automatically by the code based on the position of tails and intersection.

One of the situations that is problematic, in the method described so far, is when the tail of an object overlaps with the section of another one. In this situation, the method does an additional separation and stitching step to overcome the issue. This feature is first detected based on the fact that the number of tails of the objects is odd and they share tails. After that, all the sections are separated from intersections and then stitched together based on their orientation similar to the method described before.

## 2.4 Measurements

After separating all the objects, their length is calculated based on pixels hoping and the scale bar of the image. Individual measurement of each object by Nano1D algorithm, provides histograms for lengths of objects too. Moreover, the number of pixels in the image before skeletonization is summed to find the overall surface section area of objects. Now, the average diameter of the objects can also be calculated by dividing the area calculated to the length achieved by Nano1D algorithm.

Contrary to length measurement, this data is only an average of all objects and does not provide diameter of the objects individually, yet it can be helpful in studying one-dimensional nanoparticles.

## 3 Results and Discussion

### 3.1 Analysis of Ag nanowires

SEM image of Ag nanowires and the result of image preprocessing after skeletonization is shown in Figure. 1. As it is shown in Figure. 1b, c, both Lee and Zhang algorithm can skeletonize the nanowires effectively, while Zhang method can produce more smooth morphology at intersections. Although Lee's algorithm can skeletonize 3D objects too, in our case, Zhang's algorithm seems slightly more accurate because of producing more orthogonal hopping rather than diagonal ones. It should be noted that diagonal hopping is calculated as $\sqrt{2} \times$pixel in length calculation. Another parameter regarding different skeletonization methods is their result for tiny tails. As it is shown with green rectangles in Figure. 1a- c, Lee method ignores these tiny tails, while Zhang method detects them. Overall, comparison of these two methods showed that the algorithm benefits from around 5% higher accuracy with Zhang method with the cost of about 2% increase in computation time.

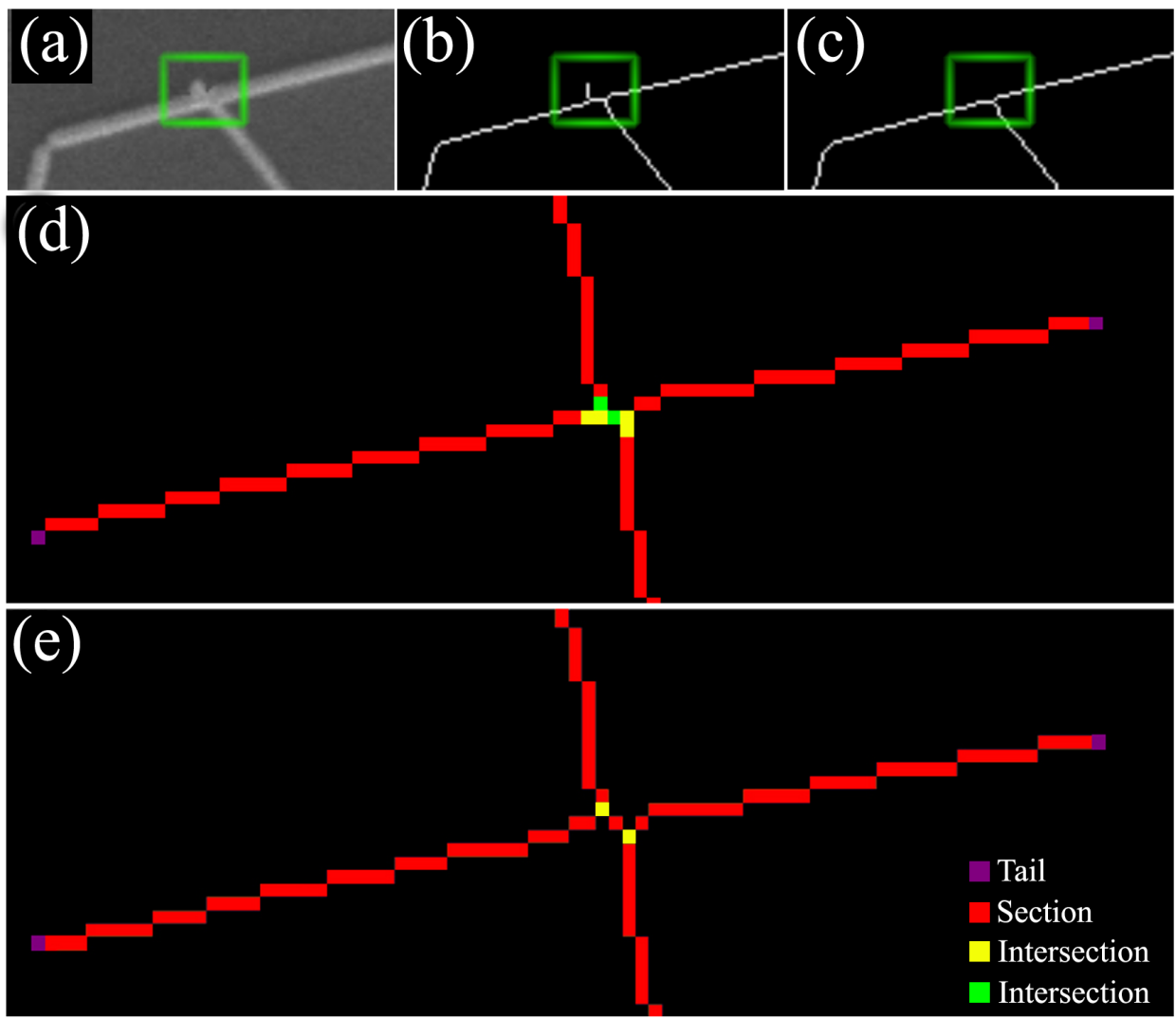

Figure. 1. (a) SEM picture of tiny tail of a nanowire, (b) the same tail skeletonized by Zhang and (c) Lee algorithm; image labeled with the nearest neighbor number of each pixel ($N(x, y)$) skeletonized based on (d) Zhang and (e) Lee algorithm

To find the tails and intersections, the nearest neighbor number is calculated based on Nano1D algorithm and the image is labeled based on that. The result of pixel labeling is shown in Figure. 1d, e. As can be observed, tails are identified with $N(x, y) =1$, intersections with $N(x, y) > 2$ and sections with $N(x, y) =2$. It should be noted that neighboring number of intersections can be 3 or 4. Skeletonization using Zhang algorithm tends to produce pixels with $N(x, y) =3$ and 4 at intersections, while Lee algorithm results in $N(x, y) = 3$. Diagonal hops produced by Lee algorithm can be problematic in proper orienteering by the algorithm, while Zhang algorithm needs more time to process more intersection pixels. This characteristic of Zhang algorithm is solved by Nano1D algorithm, which removes some unnecessary pixels and wipes out $N(x, y) = 4$. Although the last process slows down the algorithm slightly, it makes the decision making based on orientations easier and less prone to mistakes.

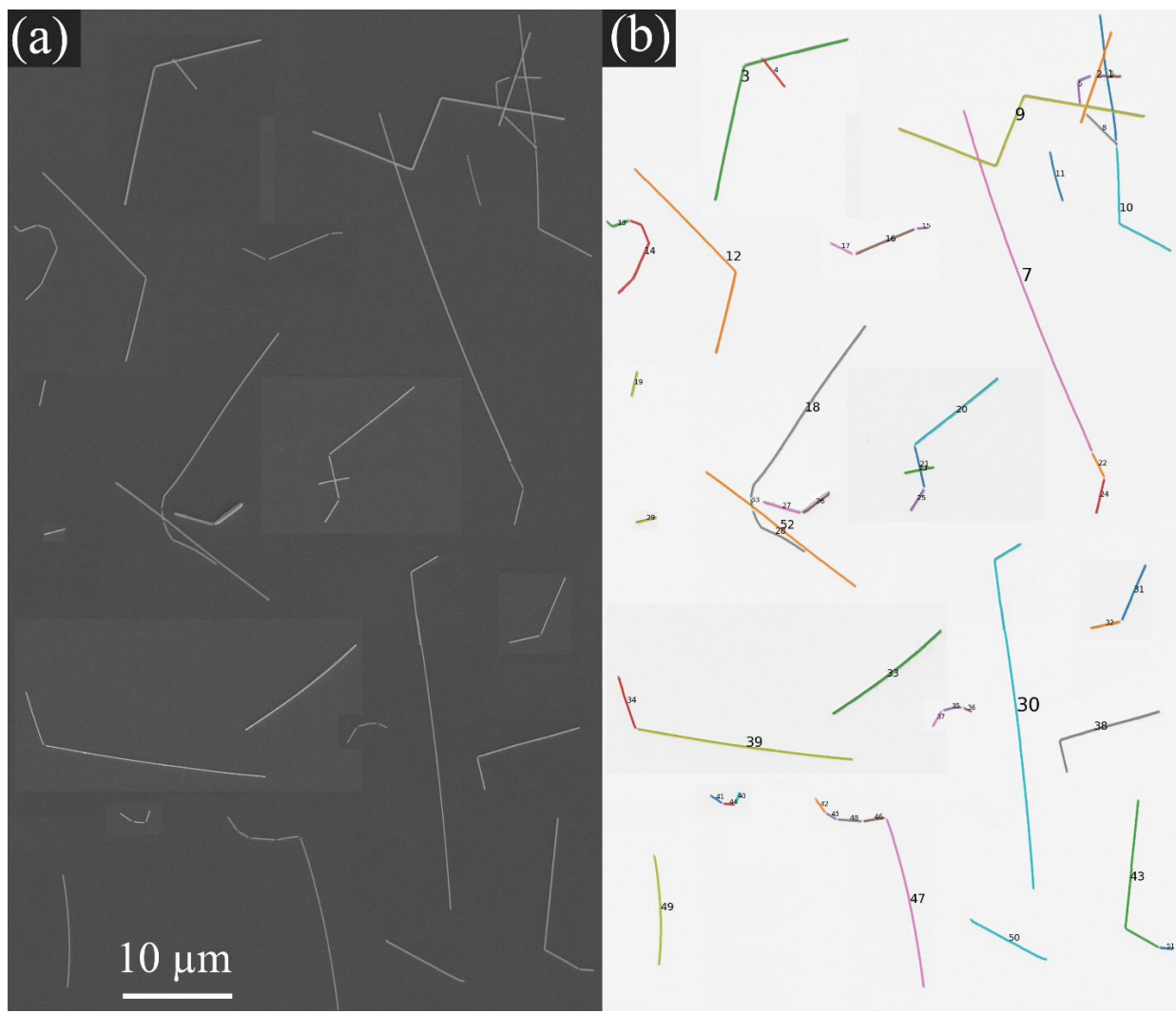

Figure. 2. (a) A tile of SEM images of Ag nanowires with 1500X magnification, (b) the same image analyzed by Nano1D.

The segmentation of overlapping nanowires by Nano1D code is shown in Figure. 2-4 and the data is shown in Figure. 5. As it is observed in Figure. 2, nanowires have been segmented regardless of their size, number, or orientation. All objects have been assigned a number by the code and labeled in their middle. Besides, small discontinuities in the objects have been detected correctly. These breaks can be more common in intersections as they can be seen in the upper right part of the images. On the upper left of the image, one tail of the nanowire number 4 is overlapped with nanowire number 3. This special situation has been dealt with by the method described in the last paragraph of section 2.3.

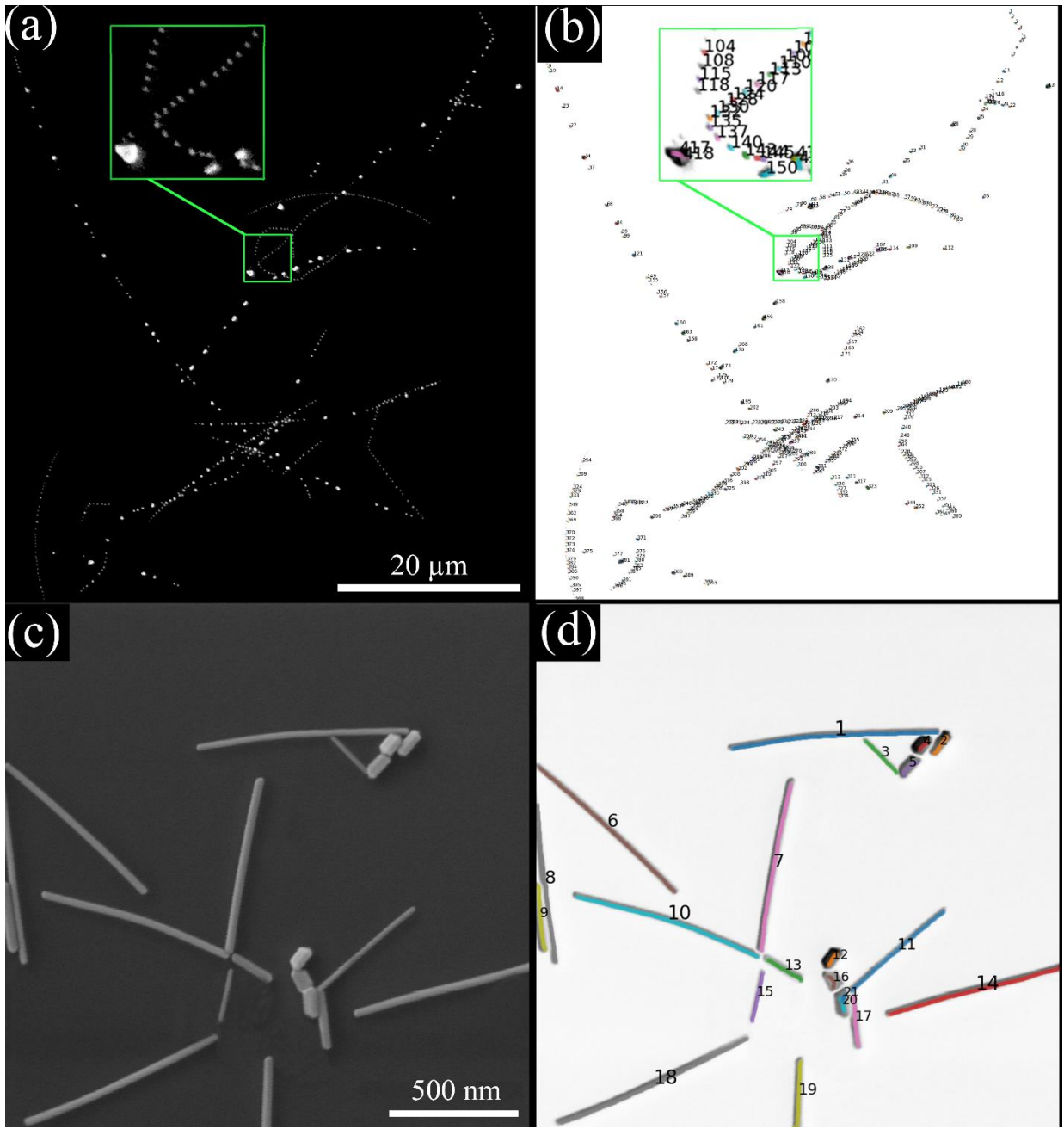

Figure. 3. (a) SEM and (b) analyzed image of Ag nanowires fragmented by thermal annealing, (c) SEM and (d) analyzed image of Au nanowires.

To evaluate the flexibility of the code in different SEM images, other images for fragmented Ag nanowires and Au nanowires were also analyzed and the results have been shown in Figure. 3. In the case of fragmented Ag nanowires (Figure. 3a), the objects have changed from 1D nanowires to small nanoparticles. Au nanowires in Figure. 3c has also developed a thick morphology in fragmented areas. In this situation, when the objects slip towards 2D realm, they can still be analyzed, but it should be noted that after skeletonization, they develop a different morphology, and the results might not always be suitable for 2D analysis. Figure. 4 also shows that another SEM image of Ag nanowires with different diameters, magnification and imaging conditions has been analyzed by the code. As it can be seen, small dusts in the image are ignored and internal noise of the objects has been resolved in the preprocessing step. Otherwise, nanowires would have seemed like 2D objects with two separate walls and an internal part.

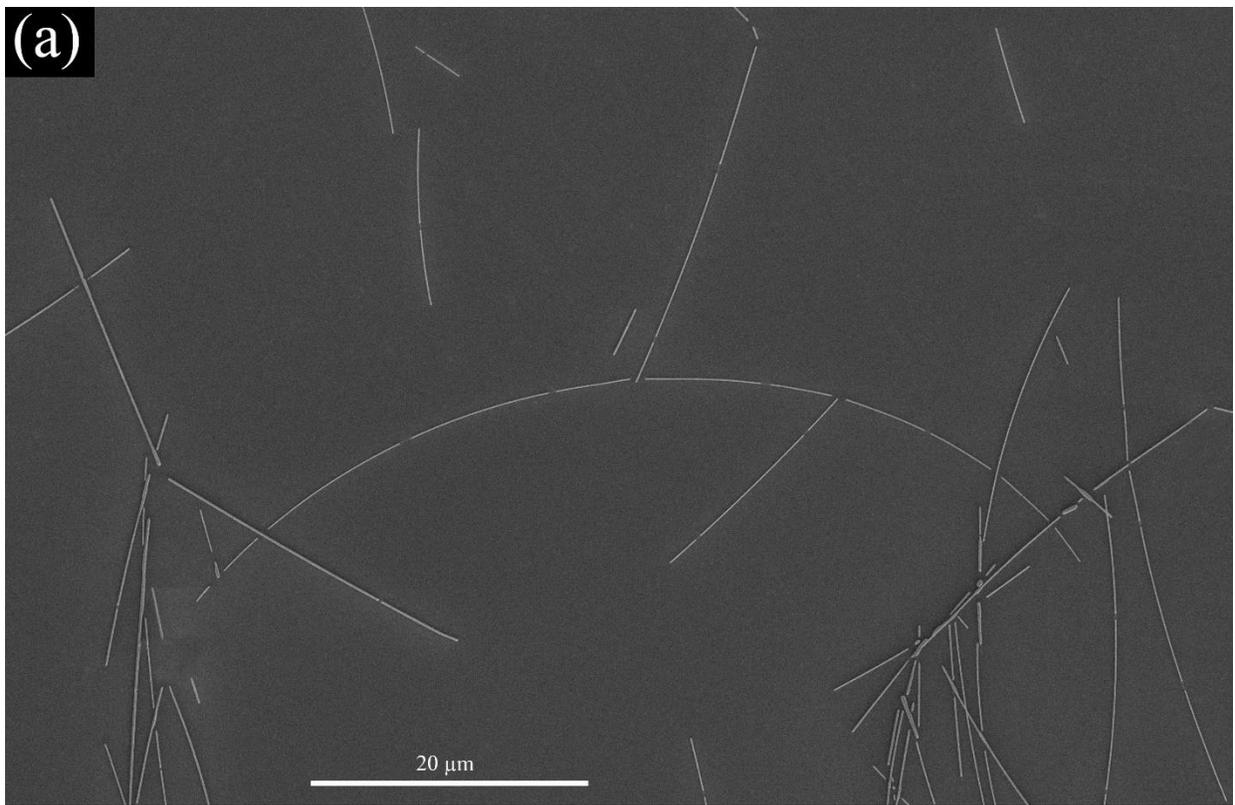
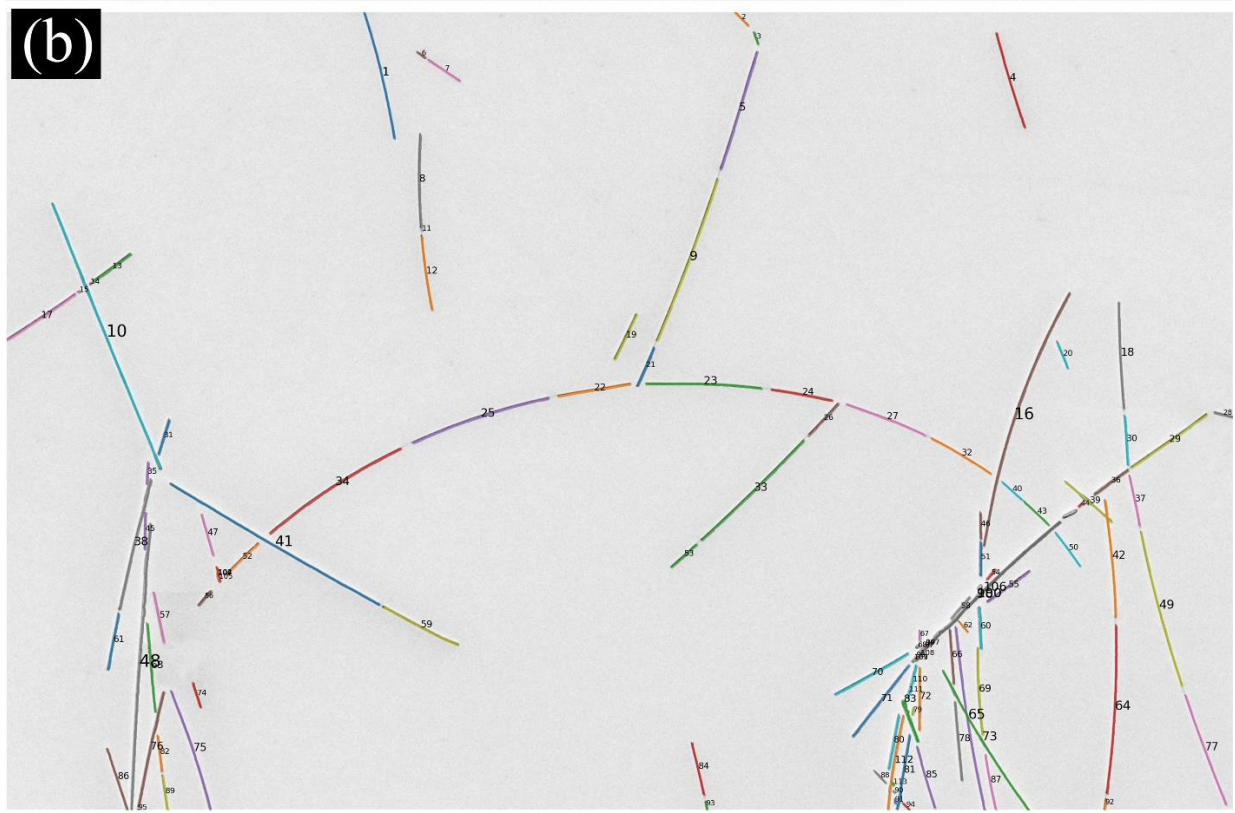

Figure. 4. (a) Ag nanowires fragmented by thermal annealing and (b) their analyzed image.

The histogram data for the last four images provided by the code have been plotted in Figure. 5. As seen, the code can perform on objects with different lengths from nanometer to tens of micrometers, and diameters from tens to hundreds of nanometers. The number of objects analyzed in one image can reach hundreds of nanoparticles. When deciding about evolution of nanowires over time, these histograms can be highly informative, because it can give hints of the effect of their length on their stability and other time-dependent parameters. Additional data like average, mode, standard deviation, and statistical error for the calculations is also provided automatically by the program.

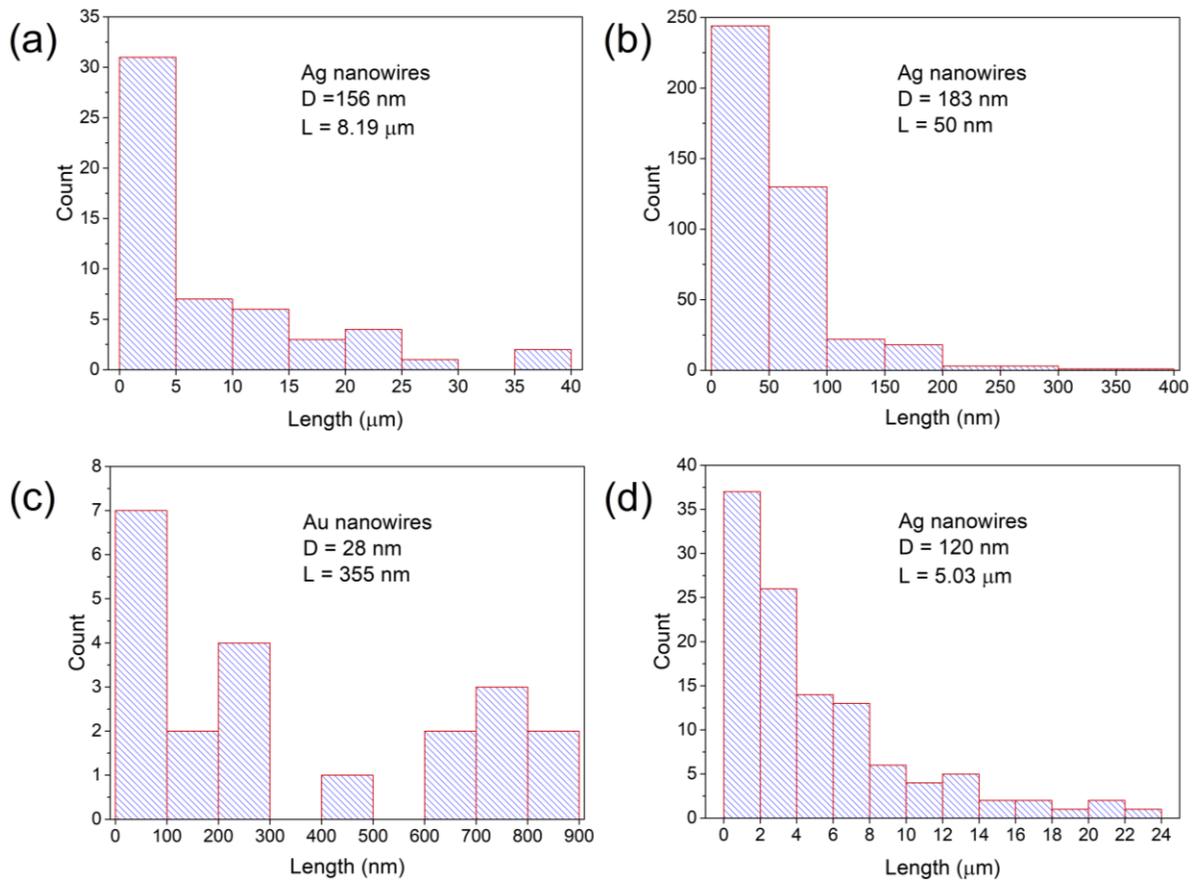

Figure. 5. Histograms of length of (a, b, d) Ag and (c) Au nanowires along with average diameter (D) and length (L).

3.2 Accuracy

Accuracy is the main strength of Nano1D, so that it barely segments objects incorrectly. The result of analysis for 10 images with 504 instances is shown in Figure. 6. As it is depicted in Figure. 6a, 7 out of 10 images have been analyzed with 100% accuracy and the remaining 3 images have also been segmented with around 96-97% accuracy. This gives the average accuracy of 99.21% over all the instances. This higher accuracy makes Nano1D much more reliable than peer ML models which barely reach 90% accuracy [1,11,14]. The small error in our algorithm happens when the tails of nanowires are so close together. As it is shown in Figure. 6b, inside the green rectangular, the algorithm has not separated two instances and detected them as one object. This mistake happens in skeletonization step and can be simply overcome by increasing the brightness threshold, but this increase may lead to fragmentation of darker nanowires and should be done with caution.

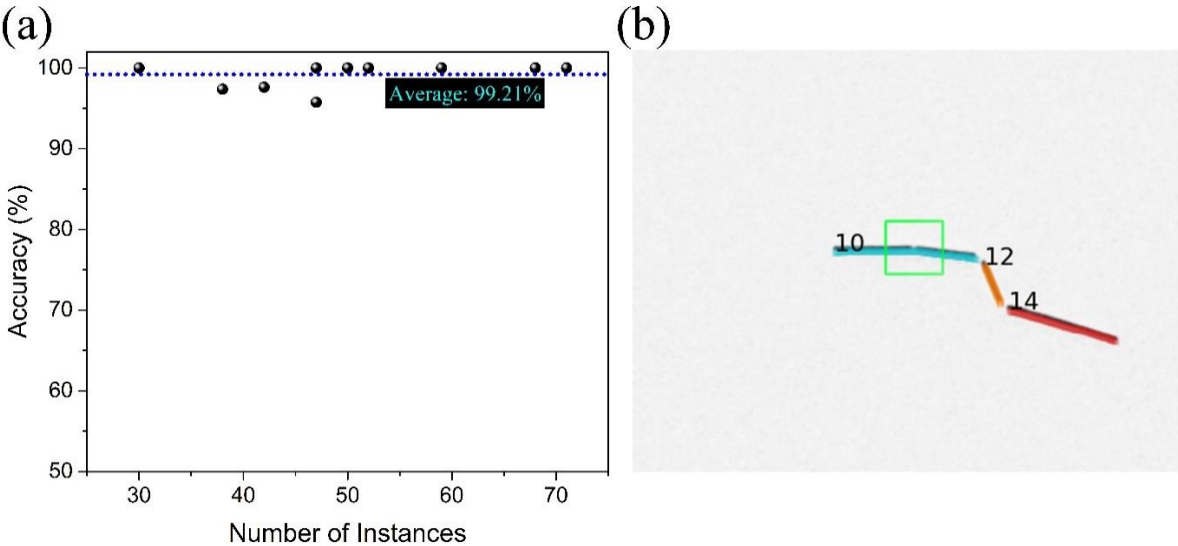

Figure. 6. (a) Segmentation accuracy of Nano1D in 10 images with 504 instances, (b) an example of the origin of inaccuracy.

While length measurements are more than 99% accurate, diameters reported by the code are averaged and suffer from a simplification that all the sections are rectangular in nature. This hypothesis can be unrealistic and make diameter values less reliable compared to length measurements. Another situation that can undermine the performance and accuracy of the code is that some objects grow thick towards becoming the edges. In this condition, the skeletonization algorithm produces closed contours that the code is unable to find the preferred orientation on intersections. Although the code escapes automatically in such undecidable situations, it is better that the inner dark colors become brightened manually or the binarization threshold be reduced to include inner morphology and avoid closed contours.

3.3 Speed

The overall processing time of Nano1D for a SEM high-resolution image with dimension of 6144×4415 pixels is around 15 seconds on just one core. Although this time is higher than detection time of ML models [2], the accuracy of our model, and multilayer data provided, makes it a more reliable asset. Furthermore, it should be noted that most ML models need time-consuming human efforts for labeling and are usually unable [1,11,15] or inaccurate in instance segmentation. Caporali model [14] which can segment overlapping instances, does this job by a combination of ML and computational methods, which make the method slow, not completely autonomous, and dependent of successive human effort and intervention.

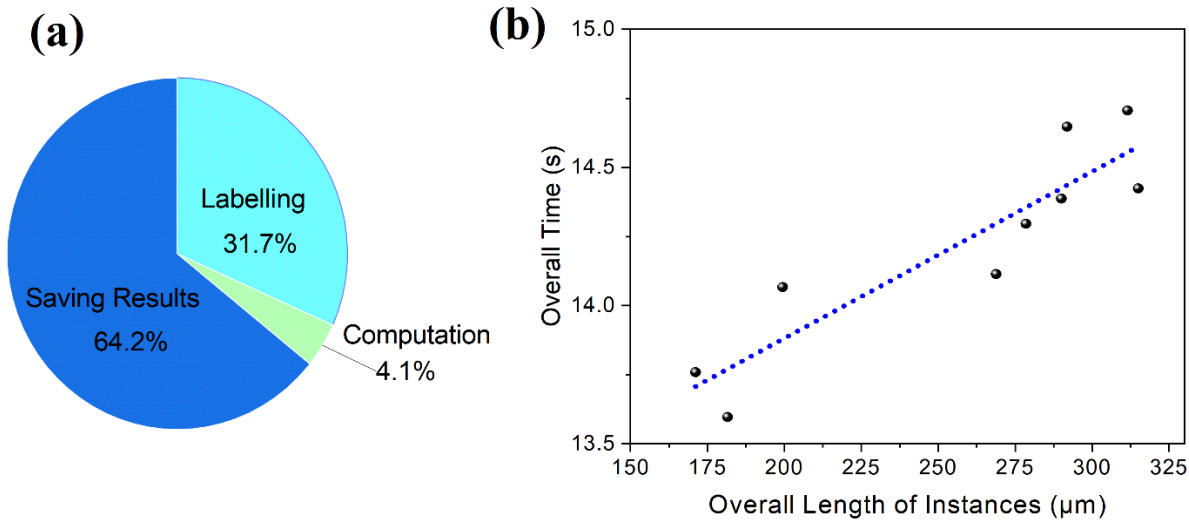

Figure. 7. (a) Proportion of labelling, computation and writing time, (b) processing time with respect to overall length of nanowires.

As it is shown in Figure. 7a, about one-third of the processing time is spent over preprocessing and labelling the image (as in Figure. 3d) and the rest is mainly used for saving the results, including neighboring number and labelled pictures and data files. The real time for segmenting the objects is about 4% of the entire time. This time is about half a second, which is comparable to ML models. Besides, it was found that analysis time is more dependent on overall length of instances rather than their number. As it is shown in Figure. 7b, doubling the overall length increases the processing time by around 10%.

3.4 Application

The Nano1D model can be used to analyze particles such as nanowires, nanotubes, nanorods, and other 1D objects in other fields. The application of the model is not only limited to 1D objects, because it can skeletonize every object (2D or 3D) and measure their 1D characteristics. Additionally, some 1D features at a macro scale, such as cracks and microstructure defects like dislocations, can also be analyzed. While detecting these features, the module for separating overlapping objects can be turned off based on the feature requirements. Considering that the overall time for analysis of one image is in order of seconds, this method can be used for mass analysis of objects and their evolution over time, e.g., growth, fragmentation, thickening, etc. Another important application of the model is for instance segmentation and data labeling for using as input data for machine learning training. The latter process is usually overwhelming to do manually, because it needs huge effort and a lot of accuracy and consideration.

## 3.5 Limitations

The main limitation of Nano1D is that it detects only one class and therefore cannot distinguish between several types of objects in each run. However, one may use multiple runs to segment different objects with different brightnesses. This limitation can be problematic when we have an image with dust particles. Although small specks of dust are removed automatically by the algorithm, bigger bright dirt should be removed manually by the user. Another area of concern is that the scale bar of microscopy image should be measured and converted to the number of pixels. These limitations can be managed in ML methods. This allows the opportunity to add an ML module to pre-process the images in the next versions. The one-dimensional approach adopted here also limits the analysis of 2D nanoparticles. A machine-learning model for analysis of 2D nanoparticles can be an undeveloped area for further research.

## 4 Conclusions

Segmentation task is a decisive step not only for measurement and analysis purposes, but also for annotation and data labelling for machine learning input. In this research, a program for segmentation and analysis of one-dimensional objects has been developed. Nano1D has been tested on Ag and Au nanowires with different lengths, diameters and population densities and showed an accuracy of about 99% in the segmentation of one-dimensional objects. The algorithm could also segment overlapping objects successfully and label them as separate objects; and the dimension and orientation of the objects does not undermine the function of the algorithm. This included instances with intersections and trisections with one tail on section of another. Regarding speed, processing time was around 15 seconds for a large 27-megapixel image, while only half a second is spent over the real segmentation process and the rest is dedicated to pre-processing and saving data. Increased density of instances also added only a few percent to analysis time. The algorithm could also analyze thicker objects, if colors of internal parts of the object are not different from the edges. The model has been shown to be a valuable tool for mass analysis of one-dimensional features in microscopy images. This includes different processes including growth, diffusion, crack propagation, dislocations, etc.

## *Data Availability*

The source code is available on:

https://gitlab.com/matter_lab/nano1d

An executable portable version of the code is also available on:

https://gitlab.ut.ee/ehsan.moradpur.tari/Nano1D/-/blob/master/Install-Nano1D.rar